\documentclass[useAMS,usenatbib]{mn2e}

\usepackage{epsfig}
\usepackage{longtable}
\usepackage{times}

\usepackage{amsmath}
\usepackage{amssymb}
\def\XMM{{\em XMM--Newton}}
\def\ROSAT{{\em ROSAT}}
\def\RXTE{{\em RXTE}}

\def\Swift{{\em Swift}}
\def\ASCA{{\em ASCA}}

\def\pn{{\em pn}}

\def\pn{{\em pn}}

\def\smc{SMC X-2}
\def\rxj{RX J0059.2-7138}
\def\approxgt{\mathrel{\hbox{\rlap{\lower.55ex \hbox {$\sim$}}
        \kern-.3em \raise.4ex \hbox{$>$}}}}
\def\approxlt{\mathrel{\hbox{\rlap{\lower.55ex \hbox {$\sim$}}
        \kern-.3em \raise.4ex \hbox{$<$}}}}

\def\flux {\mbox{erg cm$^{-2}$ s$^{-1}$}}
\def\lum {\mbox{erg s$^{-1}$}}
\def\nh {$N_{\rm H}$}
\def\ltsima{$\; \buildrel < \over \sim \;$}
\def\lsim{\lower.5ex\hbox{\ltsima}}
\def\gtsima{$\; \buildrel > \over \sim \;$}
\def\gsim{\lower.5ex\hbox{\gtsima}}

\def\nh {N${\rm _H}$}

\def\fph {ph cm$^{-2}$ s$^{-1}$}
\def\hcm {\hbox {\ifmmode $ atom cm$^{-2}\else atom cm$^{-2}$\fi}}

\def\arcsec {\hbox{$^{\prime\prime}$}}
\def\chisqnu {$\chi^{2}_{\nu}$}

\newcommand{\be}{\begin{equation}}

\newcommand{\ee}{\end{equation}}

\title[Spectral analysis of \smc]{Spectral analysis of \smc\ during its 2015 outburst}
\vspace{-0.5 cm}
\author[La Palombara et al.]{N.~La Palombara$^{1}$\thanks{E-mail: nicola@iasf-milano.inaf.it}, L.~Sidoli$^{1}$, F.~Pintore$^{1}$, P.~Esposito$^{1}$, S. Mereghetti$^{1}$ and A.~Tiengo$^{1,2,3}$ \\
$^{1}$INAF, Istituto di Astrofisica Spaziale e Fisica Cosmica, via E.\ Bassini 15,   I-20133 Milano,  Italy   \\
$^{2}$IUSS, Istituto Universitario di Studi Superiori, piazza della Vittoria 15,  I-27100 Pavia, Italy \\
$^{3}$INFN, Sezione di Pavia, via A. Bassi 6, I-27100 Pavia, Italy \\
}

\begin{document}
\vspace{-0.5 cm}

\date{Accepted: 2016 February 8. Received: 2015 December 23.}
\vspace{-0.5 cm}

\pagerange{\pageref{firstpage}--\pageref{lastpage}} \pubyear{2016}
\vspace{-0.5 cm}

\maketitle
\vspace{-0.5 cm}

\label{firstpage}
\vspace{-0.5 cm}

\begin{abstract}
We report on the results of \XMM\ and \Swift\ observations of \smc\ during its last outburst in 2015 October, the first one since 2000. The source reached a very high luminosity ($L \sim 10^{38}$ \lum), which allowed us to perform a detailed analysis of its timing and spectral properties. We obtained a pulse period $P_{\rm spin}$ = 2.372267(5) s and a characterization of the pulse profile also at low energies. The main spectral component is a hard ($\Gamma \simeq$ 0) power-law model with an exponential cut-off, but at low energies we detected also a soft (with kT $\simeq$ 0.15 keV) thermal component. Several emission lines are present in the spectrum. Their identification  with the transition lines of highly ionized N, O, Ne, Si, and Fe suggests the presence of photoionized matter around the accreting source.
\end{abstract}

\vspace{-0.5 cm}

        \section{Introduction\label{intro}}

\smc\ is one of the first pulsars discovered in the Small Magellanic Cloud (SMC). It was discovered with \textit{SAS 3} in 1977 \citep{Li+77,Clark+78}, at a luminosity $L_{\rm (2-11~keV)}$ = 8.4 $\times 10^{37}$ \lum, and the lack of detection in an observation performed one month later showed its transient nature \citep{Clark+79}. Between 1991 and 1992 it was visible only in one of two \ROSAT\ observations performed six months apart \citep{KahabkaPietsch96,Sasaki+00}, implying a dynamic range $> 6\times 10^{3}$. A second outburst was observed in 2000 with \RXTE, when the spin period of 2.37 s was measured \citep{Corbet+01}. The pulse period was confirmed by a follow-up observation performed by \ASCA, which identified the pulsar discovered with \RXTE\ with \smc\ \citep{Torii+00,Yokogawa+01}. 

The optical counterpart, originally identified by \citet{Crampton+78}, was later resolved into two different stars of early spectral type, separated by $\sim$ 2.5\arcsec \citep{Schmidtke+06}. Both stars were monitored by the Optical Gravitational Lensing Experiment (OGLE, \citealt{Udalski+03}). The OGLE-III data revealed that the 
southern, fainter star is almost constant, while the northern star has a periodic variability (by up to 1 mag) 
with $P$ = 18.62 $\pm$ 0.02 d \citep{Schurch+11}. \RXTE\ measured a periodic modulation of the pulse period at $P$ = 18.38 $\pm$ 0.02 d \citep{Townsend+11}. This strongly suggests that the northern star, which is an O9.5 III-V emission-line star \citep{McBride+08}, is the true counterpart and that the observed periodicity is the orbital period of the binary system.

After more than 15 years, in September 2015 \smc\ showed a new outburst \citep{Negoro+15,Kennea+15}, during which it reached a very high luminosity ($L \sim 10^{38}$ \lum). The source was monitored with \Swift\ and we obtained a follow-up ToO observation with \XMM. In this paper we report on the results obtained
with these observations.
\vspace{-0.5 cm}

 	 \section{Observations and Data Reduction}
         \label{data}

\XMM\ observed \smc~between 2015 October 8 and 9, for a total exposure time of 30 ks. The three EPIC cameras, i.e.~one \pn\ \citep{Struder+01} and two MOS \citep{Turner+01}, were all operated in \textit{Small Window} mode, with time resolution of 5.7 ms and 0.3 s for the \pn\ and the MOS cameras, respectively; for all cameras the Thin filter was used. The Reflection Grating Spectrometer (RGS) was operated in \textit{spectroscopy} mode \citep{denHerder+01}.

We used version 14 of the \XMM~{\em Science Analysis System} (\textsc{sas}) to process the event files. After the standard pipeline processing, we searched for possible intervals of high instrumental background. The last $\simeq$ 2 ks of the observation were affected by a high background level and rejected. Taking into account also the dead time (29 \% and 2.5 \% for \pn\ and MOS, respectively), the effective exposure time was 19.7 ks for \pn\ camera and $\simeq$ 27 ks for the two MOS cameras.

For the analysis of the EPIC data we selected events with pattern in the range 0--4 (mono-- and bi--pixel events) for the \pn\ camera and 0--12 (from 1-- to 4--pixel events) for the two MOS. Due to the very high count rate of the source, both the \pn\ and MOS data were significantly affected by photon pile-up. Therefore, we selected events from an annular region around the source position, ignoring those from the inner circular area: for the \pn\ camera we selected events between 10 and 45'' from the source position, while for both MOS cameras we considered an extraction region between 20 and 40''. In both cases we performed a fit of the radial profile with a King function to define the inner radius; instead the outer radius was limited by the CCD edge or dark columns. For each camera, background events were selected from circular regions offset from the target position.

All EPIC and RGS spectra were fitted using \textsc{xspec} 12.7.0. In the following, all spectral uncertainties and upper limits are given at the 90 \% confidence level for one interesting parameter.

For the timing analysis we considered also 74 observations performed by \Swift/XRT in \textit{Windowed Timing} (WT) mode. We carried out their data reduction using the standard \textsc{xrtpipeline}. Source events were then extracted from a circular region of 47\arcsec\ radius around the source position.
\vspace{-0.75 cm}

 	 \section{Timing analysis}
         \label{timing}

\begin{figure}
\begin{center}
\includegraphics[height=6.25cm,angle=0]{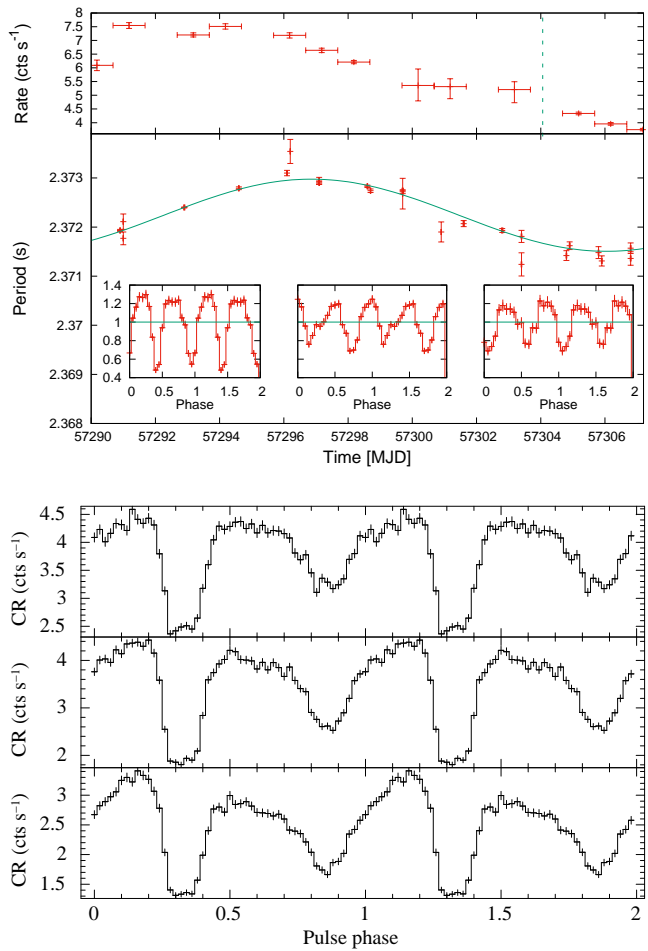}
\includegraphics[width=6cm,angle=-90]{source_pn_10-45_bar_BaryBin_clean_3E_50bins_fef.ps}
\caption{\textit{Upper panel}) Flux and spin-period evolution of \smc\ during  the 2015 outburst, as observed by \Swift/XRT; in the insets, we show three pulse profiles at different stages of the outburst (around MJD 57291, 57299 and 57303). The vertical dashed line corresponds to the epoch of the \XMM\ observation. \textit{Lower panel}: EPIC \pn\ pulse profile of \smc\ in the 0.15--2 keV (top), 2--5 keV (middle), and 5--12 (bottom) energy range.}
\label{lc}
\end{center}
\vspace{-0.75 cm}
\end{figure}

For each \Swift/XRT event file, the photon times of arrival were reported at the solar system barycenter applying the {\sc barycorr} tool; then the source pulsation period during each observation was measured by fitting the peak in the distribution of the Rayleigh test statistics as a function of the trial period. In Fig.~\ref{lc} (upper panel) we present the flux and spin period evolution along the outburst; a modulation of the pulse period induced by the orbital motion is clearly visible. We fitted the pulse period values with a constant plus a sinusoid obtaining a period of $18.38\pm0.96$ days and an amplitude corresponding to a projected semi-axis $A_x$sin($i$) = 78 $\pm$ 3 light-seconds. This is consistent within 2$\sigma$ with the previously reported value of 73.7 $\pm$ 0.9 light-seconds \citep{Townsend+11}.
Although poorly constrained, we were also able to estimate the time of the passage at the ascending node, $T^* = 57297 \pm 6$ MJD. 
The constant term in the fit provided $P_{\rm spin}$ = 2.37224(2) s.

Adopting these parameters, we corrected the photon times of arrival for the orbital motion in order to search for possible variations in the spin period. However, no robust hints of spin-up or spin-down were found because it was not possible to phase-connect the different observations due to the large uncertainties in the individual period measurements and to the time varying pulse profile.

For the timing analysis of the \XMM\ observation  we used the \pn\ data in the energy range 0.15-12 keV. The event arrival times were converted to the solar system barycenter (with the \textsc{sas} tool \textsc{barycenter})  and to the binary system barycenter, based on the results obtained with \Swift. We then measured the pulse period by a standard phase-fitting technique, obtaining a best-fit value $P_{\rm spin}$ = 2.372267(5) s and \textbar $\dot P$ \textbar $< 7\times10^{-9}$ s s$^{-1}$ (3$\sigma$ c.l.).

In the lower panel of Fig.~\ref{lc} we show the folded light curves in the energy ranges 0.15--2, 2--5, and 5--12 keV. The shape of the pulse profile is similar in the three ranges. It shows two broad peaks, of comparable width (0.3-0.4 in phase) and amplitude, separated by a primary and a secondary minimum. The pulse profile is smooth; around the primary minimum the count rate (CR) increase/decrease is very fast, while it is slower around the secondary minimum. The pulsed fraction, defined as (CR$_{\rm max}$ -- CR$_{\rm min}$)/(2 $\times$ CR$_{\rm average}$), is between $\sim$ 30 (for the soft range) and 40 \% (for the hard range). Finally, we have verified that also in narrower energy ranges the pulse profile is characterized by the same properties; in particular, also below 0.5 keV the pulse profiles shows two different peaks (although with a pulsed fraction of $\sim$ 10 \% only).


We accumulated also background--subtracted light curves over the whole \XMM\ observation; then, we used the (\textsc{sas}) tool \textsc{epiclccorr} to correct each light curve for the extraction region. We found that, for both \pn\ and MOS cameras, the total count rate in the full range 0.15--12 keV was $\sim$ 43 and 15 c s$^{-1}$, respectively: these values are well above the limit to avoid deteriorated photon pile-up\footnote{http://xmm.esac.esa.int/external/xmm\_user\_support/documentation/uhb/ epicmode.html}, thus confirming the need to reject events from the inner circular region. All the light curves show a significant flux variability (up to $\sim$ 20 \% on time scales of 100 s), but without any long term increasing/decreasing trend  over the whole observation; moreover, although also the ratio between the fluxes in different energy ranges is variable, there is no evidence of a long-term spectral variation.
\vspace{-0.5 cm}

  	\section{Spectral analysis}

Since we found no evidence for long-term intensity or spectral variability of \smc\ along the \XMM\ observation, we performed a time-averaged analysis of the EPIC spectrum. The response matrices and ancillary files for the source and background spectra were generated using the \textsc{sas} tasks \textsc{rmfgen} and \textsc{arfgen}, and the spectral analysis was performed in the energy range 0.3--12 keV, fitting the three EPIC spectra simultaneously; they were rebinned with a minimum of 30 counts per bin. To account for uncertainties in instrumental responses, we introduced for the MOS detectors normalization factors relative to the \pn\ camera (1.025 $\pm$ 0.009 for MOS1 and 1.039 $\pm$ 0.009 for MOS2). We adopted the interstellar abundances of \citet{WilmsAllenMcCray00} and photoelectric absorption cross-sections of \citet{BalucinskaChurchMcCammon92}, using the absorption model \textsc{phabs} in \textsc{xspec}.

\begin{figure}
\includegraphics[height=8.5cm,angle=-90]{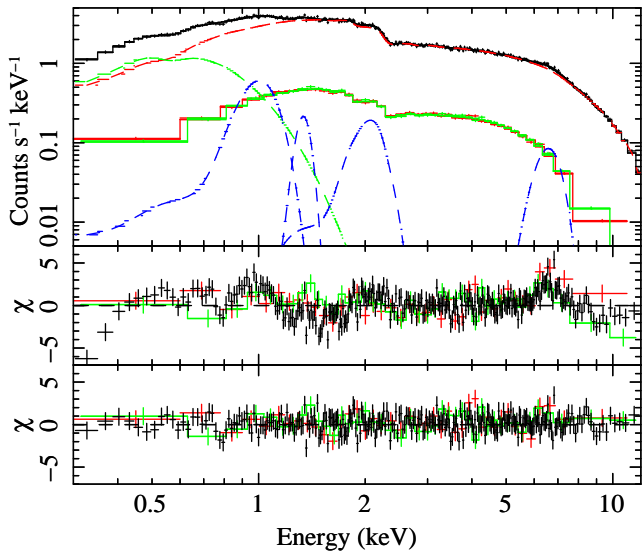}
\caption{Time-averaged spectrum of \smc; \pn, MOS1, and MOS2 data are reported in black, red, and green, respectively. \textit{Upper panel:} superposition of the EPIC spectra with (for the \pn\ spectrum only) the best-fit \textsc{cutoffpl+bb} model (red and green dashed lines) plus the additional gaussian components (blue dashed lines). \textit{Middle panel:} data-model residuals in the case of the fit with a simple \textsc{cutoffpl} model. \textit{Lower panel:} data-model residuals in the case of the best-fit model.}
\label{spectrum}
\vspace{-0.5 cm}
\end{figure}

The source spectrum shows a clear high-energy cut-off above $\sim$ 7 keV (Fig.~\ref{spectrum}, upper panel), therefore we fit it with an absorbed cut-off power-law model (\textsc{cutoffpl} in \textsc{xspec}), defined as $A(E) = kE^{-\Gamma}$ exp(-$E/E_{\rm cut}$). We obtained a reasonable fit (\chisqnu/d.o.f. = 1.17/2880), but the residuals show several significant features (Fig.~\ref{spectrum}, middle panel): 1) a soft excess around 0.5 keV; 2) a broad structure around $\sim$ 1.0 keV; 3) a feature at $\sim$ 2 keV; 4) another structure at $\sim$ 6.5 keV. The latter can be attributed to a blend of emission lines of Fe at different ionization levels (see the Discussion) and we described it with a Gaussian component. The feature at $\sim$ 2 keV, present only in the \pn\ data, is very likely due to residual calibration uncertainties around the Au edge; as done by other authors \citep[e.g.][]{DiazTrigo+14}, we modeled also this structure with a Gaussian and we will not discuss it further. On the other hand, a calibration/instrumental origin for the broad emission feature at $\sim$ 1 keV can be excluded, since emission features around 1 keV were clearly detected also in the RGS spectra (see below). In order to describe both this structure and the soft excess around 0.5 keV, we considered two different possibilities: 1) a blackbody (BB) component plus a Gaussian line; 2) an emission spectrum from collisionally-ionized gas (\textsc{apec} in \textsc{xspec}). In Table~\ref{epic_fit} we report the best-fit parameters obtained for both possibile deconvolutions.

In the first case, the addition of a BB component (with kT $\simeq$ 0.2 keV) and a Gaussian line at 1 keV reveals the presence of an additional emission feature at $\simeq$ 1.35 keV. Therefore, the fit of the overall spectrum requires a \textsc{cutoffpl+bb} model to describe the spectral continuum, plus four additional Gaussian components to describe the various structures at 1, 1.35, 2, and 6.5 keV (Fig.~\ref{spectrum}, upper and lower panels). 

In the second case, a single \textsc{apec} component at kT $\simeq$ 1.2 keV can account for both the soft excess and the feature at 1 keV. In this way, the description of the spectral continuum with an absorbed \textsc{cutoffpl+apec} model requires only two additional Gaussian components at 2 and 6.5 keV. With this model, if in the \textsc{apec} component the metal abundance is left free to vary, its best-fit value is 0.034$^{+0.015}_{-0.014}$, i.e. well below the estimated metallicity $Z \simeq 0.2 Z_{\odot}$ for the SMC \citep{Russel&Dopita92}. We note that a good fit can be obtained also with the abundance value fixed at 0.2, with a \chisqnu\ comparable to that obtained with a free abundance.


\begin{table*}
\caption{Results of the simultaneous fit of the time-averaged spectrum of the \pn\ and MOS data. The double-component continuum consists of a cut-off power-law and either a blackbody or a thermal plasma model; in the second case both a free and a fixed solar abundance is considered. In addition, various Gaussian lines in emission are needed to account for positive residuals in the spectrum.}\label{epic_fit}
\begin{tabular}{cccc} \hline
Continuum Model					& \textsc{cutoffpl+bb}			& \textsc{cutoffpl+apec}		& \textsc{cutoffpl+apec}		\\
Parameter					& 					& (free abundance)			& (fixed abundance)			\\ \hline
\nh (10$^{21}$ cm$^{-2}$)			& 1.8$\pm$0.3				& 1.4$\pm$0.1				& 0.80$^{+0.05}_{-0.06}$		\\
$\Gamma$					& 0.12$^{+0.07}_{-0.08}$		& -0.09$\pm$0.07			& 0.11$^{+0.03}_{-0.04}$		\\
$E_{\rm cut}$ (keV)				& 6.9$^{+0.7}_{-0.6}$			& 5.7$^{+0.4}_{-0.3}$			& 7.0$^{+0.4}_{-0.3}$			\\
Flux$_{\rm CPL}$ (0.3-12 keV, $\times 10^{-10}$\flux)		& 3.24$^{+0.02}_{-0.03}$ & 3.17$^{+0.02}_{-0.03}$& 3.22$^{+0.03}_{-0.02}$				\\
$kT_{\rm BB~or~APEC}$ (keV)			& 0.135$^{+0.014}_{-0.011}$		& 1.2$\pm$0.1				& 1.22$^{+0.07}_{-0.10}$		\\
$R_{\rm BB}$ (km) or $N_{\rm APEC}$ (cm$^{-5}$)	& 320$^{+125}_{-95}$			& 1.9$^{+0.4}_{-0.3} \times 10^{-2}$	& (4.8$\pm$1.1)$\times 10^{-3}$		\\
Flux$_{\rm BB~or~APEC}$ (0.3-12 keV, $\times 10^{-12}$ \flux)	& 7.5$^{+2.5}_{-1.9}$	& 12.6$^{+1.6}_{-1.1}$& 3.9$\pm$0.9						\\
Abundance (\textsc{apec})			& -					& 0.034$^{+0.015}_{-0.014}$		& 0.2 (fix)				\\
$E_{\rm line1}$ (keV)				& 0.99$^{+0.03}_{-0.02}$		& -					& -					\\
$\sigma_{\rm line1}$ (keV)			& 0.09$^{+0.03}_{-0.02}$		& -					& -					\\
Flux$_{\rm line1}$ ($\times 10^{-4}$ \fph)	& 5.5$^{+2.6}_{-2.2}$			& -					& -					\\
EW$_{\rm line1}$ (eV)				& 52$^{+14}_{-16}$			& -					& -					\\
$E_{\rm line2}$ (keV)				& 1.32$^{+0.05}_{-0.08}$		& -					& -					\\
$\sigma_{\rm line2}$ (keV)			& 0.13$^{+0.09}_{-0.05}$		& -					& -					\\
Flux$_{\rm line2}$ ($\times 10^{-4}$ \fph)	& 3.2$^{+3.0}_{-1.6}$			& -					& -					\\
EW$_{\rm line2}$ (eV)				& 38$^{+13}_{-16}$			& -					& -					\\
$E_{\rm line3}$ (keV)				& 1.99$^{+0.05}_{-0.07}$		& 2.08$\pm$0.05				& 2.08$^{+0.05}_{-0.06}$		\\
$\sigma_{\rm line3}$ (keV)			& 0.26$^{+0.12}_{-0.08}$		& 0.21$^{+0.07}_{-0.05}$		& 0.16$^{+0.05}_{-0.04}$		\\
Flux$_{\rm line3}$ ($\times 10^{-4}$ \fph)	& 4.0$^{+3.2}_{-1.7}$			& 2.8$^{+1.1}_{-0.7}$			& 1.6$^{+0.6}_{-0.5}$			\\
EW$_{\rm line3}$ (eV)				& 58$^{+16}_{-21}$			& 41$^{+9}_{-6}$			& 24$^{+4}_{-5}$			\\
$E_{\rm line4}$ (keV)				& 6.62$\pm$0.09				& 6.62$^{+0.10}_{-0.09}$		& 6.60$\pm$0.09				\\
$\sigma_{\rm line4}$ (keV)			& 0.46$^{+0.15}_{-0.10}$		& 0.39$^{+0.11}_{-0.09}$		& 0.48$^{+0.15}_{-0.10}$		\\
Flux$_{\rm line4}$ ($\times 10^{-4}$ \fph)	& 3.7$^{+1.1}_{-0.9}$			& 2.9$^{+0.8}_{-0.7}$			& 3.9$^{+1.0}_{-0.8}$			\\
EW$_{\rm line4}$ (eV)				& 121$^{+15}_{-11}$			& 93$^{+11}_{-12}$			& 126$^{+16}_{-15}$			\\
L$_{\rm BB~or~APEC}$/L$_{\rm CPL}$ (0.01-12 keV)& 3.1 \%				& 6.0 \%				& 1.8 \%				\\
Unabsorbed flux (0.3-12 keV, $\times 10^{-10}$ \flux)		& 3.38$\pm$0.04	& 3.34$\pm$0.02	& 3.31$\pm$0.02	\\
Luminosity (0.3-12 keV, $\times 10^{38}$ \lum)			& 1.42$\pm$0.02	& 1.41$\pm$0.01	& 1.39$\pm$0.01 	\\
\chisqnu/d.o.f.					& 1.03/2866				& 1.05/2871				& 1.07/2872				\\ \hline
\end{tabular}
\end{table*}


\begin{figure}
\centering
\resizebox{\hsize}{!}{\includegraphics[angle=-90,clip=true]{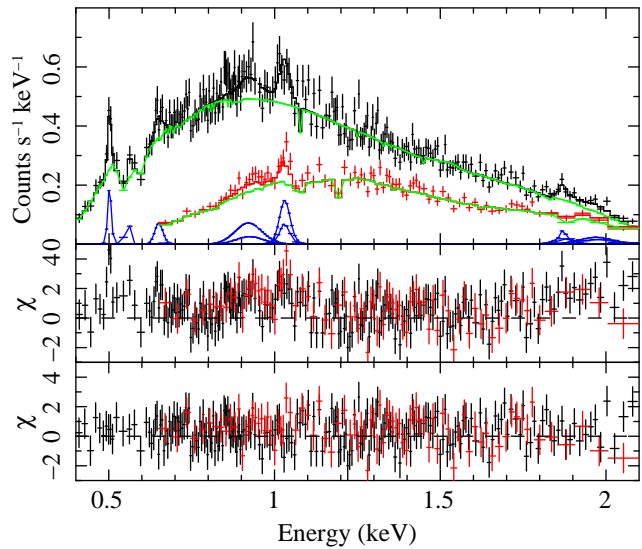}}
\caption{Combined RGS1 and RGS2 spectra for the first and the second order (\textit{black} and \textit{red} symbols, respectively); for displaying purposes we applied a graphical rebinning. \textit{Upper panel:} superposition of the spectrum with the best-fit absorbed power-law model (green line) plus Gaussian components (blues lines, Table~\ref{lines}). \textit{Middle panel:} data-model residuals in the case of the fit with a simple power-law model. \textit{Lower panel:} data-model residuals in the case of the best-fit model.}
\label{rgs_spectra}
\vspace{-1 cm}
\end{figure}

The RGS1 and RGS2 data were combined into one single grating spectrum, separately for the first and the second order spectra, using the \textsc{sas} task \textsc{rgscombine}; then, the two combined spectra were analysed in the energy range 0.4--2.1 keV. We rebinned the spectra with a minimum of 20 counts per bin.

They were fitted with an absorbed power-law model, which left several emission residuals: two broad structures at $\simeq$ 0.92 and 1.98 keV and narrow emission features at $\simeq$ 0.5, 0.55, 0.65, 1.03, and 1.87 keV. It was possibile to describe all these features with the addition of Gaussian lines; their parameters are reported in Table~\ref{lines}, while the two RGS spectra are shown in Fig.~\ref{rgs_spectra}. The lines at 0.5, 0.56, and 0.65 keV were well constrained and can be associated with N {\sc vii}, O {\sc vii} (f, forbidden line), and O {\sc viii} Ly$\alpha$ lines, respectively. The broad component at 0.92 keV can be due to either a blend of several emission lines from iron in a range of ionizations states (from Fe {\sc xviii} to Fe {\sc xx}) or a radiative recombination continuum (RRC) from O {\sc viii} - Ne {\sc ix}. The emission features at 1.03, 1.87 keV and 1.98 are probably due to Ne {\sc x}, Si {\sc xiii}, and Si {\sc xiv}, respectively. In any case, all the features reported in Table~\ref{lines} are significant at least at 3$\sigma$ confidence level.

For completeness, we performed also a simultaneous fit of the EPIC and RGS spectra. The corresponding results are fully consistent with those previously shown, since the \textsc{cutoffpl+bb} or \textsc{cutoffpl+apec} models used for the EPIC continuum can also describe the continuum component of the RGS spectra. However, we note that in both cases the RGS spectra show residuals, comparable to those reported in the middle panel of Fig.~\ref{rgs_spectra}.
\vspace{-0.5 cm}

\begin{table*}
\caption{Best-fit parameters of the lines identified in the RGS spectra (0.4-2.1 keV) of \smc.}\label{lines}
\begin{tabular}{cccccc} \hline
Observed		& Ion						& Laboratory	& $\sigma$		& Flux					& EW			\\
Energy			&						& Energy	& (eV)			& (10$^{-5}$ ph cm$^{-2}$ s$^{-1}$)	& (eV)			\\
(eV)			&						& (eV)		&			&					&			\\ \hline
501$\pm$2		& N {\sc vii}					& 500		& 4.6$^{+2.4}_{-0.9}$	& 16.6$^{+11.1}_{-7.1}$			& 10.7$^{+4.0}_{-4.2}$	\\
557$^{+5}_{-4}$		& O {\sc vii}					& 561		& 4.0$^{+6.1}_{-2.9}$	& 10.1$^{+8.2}_{-5.4}$			& 6.8$^{+4.2}_{-3.9}$	\\
650$^{+7}_{-8}$		& O {\sc viii}					& 654		& 11.6$^{+9.5}_{-4.8}$	& 8.4$^{+5.5}_{-3.8}$			& 6.4$^{+4.2}_{-4.1}$	\\
920$^{+15}_{-17}$	& Fe {\sc xviii} - Fe {\sc xx} (blended ?)	& -		& 40$^{+18}_{-21}$	& 15.9$^{+7.3}_{-5.1}$			& 15.3$^{+7.5}_{-6.8}$	\\
			& RRC from O {\sc viii} - Ne {\sc ix} (?)	&		&			&					&			\\
1031$^{+5}_{-6}$	& Ne {\sc x}					& 1022		& 13.8$^{+6.2}_{-3.6}$	& 13.0$\pm$3.2				& 13.3$^{+4.3}_{-4.8}$	\\
1872$^{+4}_{-10}$	& Si {\sc xiii}					& 1860		& $<$ 19		& 9.3$^{+3.5}_{-5.4}$			& 13.5$^{+5.2}_{-8.0}$	\\
1977$^{+26}_{-33}$	& Si {\sc xiv}					& 1979		& 43.8$^{+58.3}_{-16.7}$& 19.5$^{+15.7}_{-9.5}$			& 30.8$^{+15.3}_{-21.9}$\\ \hline
\end{tabular}
\end{table*}

	      \section{Discussion}\label{sec:discussion}

The September 2015  outburst was the first one detected from \smc\ since that of 2000. Assuming a distance of 61 kpc \citep{Hilditch+05}, the unabsorbed flux $f_{\rm X} = 3.4 \times 10^{-10}$ \flux observed by \XMM\ (in the energy range 0.3-12 keV) implies a luminosity $L_{\rm X} = 1.4 \times 10^{38}$ \lum. This is comparable to the highest luminosities previously observed for this source, with \textit{SAS 3} in 1977 (Clark et al. 1978) and with \ROSAT/PSPC in 1991 \citep{KahabkaPietsch96,Sasaki+00}; it is also comparable with the peak flux detected by \RXTE/ASM in 2000 \citep{Corbet+01}. On the other hand, it is higher than the luminosities observed in 2000 with \RXTE/PCU ($L \sim 3 \times 10^{37}$($d$/65 kpc)$^2$ \lum, \citealt{Corbet+01}) and with \ASCA\ ($L \sim 4 \times 10^{36}$($d$/65 kpc)$^2$ \lum, \citealt{Yokogawa+01}).

The spin period measured by \XMM, corrected for the orbital motion, is $P_{\rm spin, 2015}$ = 2.372267(5) s, which compared to that measured by \RXTE\ in 2000 ($P_{\rm spin, 2000}$ = 2.37194(1) s, \citealt{Townsend+11}) implies an average spin-down rate of  $\dot P_{\rm spin}$ = (6.6$\pm$0.2)$\times 10^{-13}$ s s$^{-1}$ during the the $\sim$ 15 years between the two outbursts. 

The pulse profile is characterized by a double peak, not only at high energies as already observed by \RXTE\ \citep{Corbet+01}, but also at very low energies. This is at odds with the results obtained with \ASCA, which detected only a single, broad peak below 2 keV, although the count statistics was high \citep{Yokogawa+01}. This difference could be related to the factor $\sim$ 25 higher  luminosity during the \XMM\ observation.
On the other hand, the pulsed fraction measured at high energies ($\sim$ 40 \%) is comparable to that observed with \RXTE\ in 2000. From this point of view, it is interesting to compare our results on \smc\ with those obtained for \rxj, another transient pulsar in the SMC with a similar pulse period (2.76 s). This source was observed during two different outbursts in 1993 \citep{Kohno+00} and in 2014 \citep{Sidoli+15}, at a luminosity of $\sim 2 \times 10^{38}$ and $\sim 7 \times 10^{37}$ \lum, respectively. In the first outburst its  luminosity and pulsed fraction  (37 \%) were similar to those of \smc, while in 2014 both the luminosity and the pulsed fraction (9 \%) were much lower.

The \XMM\ observation of \smc\ has provided the detection of previously unknown spectral features. Although the EPIC spectrum is dominated by a hard ($\Gamma \sim 0$) cut-off power law, its fit requires the addition of a thermal component, either a soft blackbody (kT $\sim$ 0.1 keV) or a hot thermal plasma model (kT $\sim$ 1 keV); in both cases the soft component contributes for only a few \% to the total  luminosity, but it is the dominant component below 0.5 keV and its addition improves significantly the spectral fit. The size of the thermal component implies emission up to large distances from the NS. Since we observed \smc\ with a very high luminosity, based on the emission models proposed by \citet{Hickox+04} the observed BB emission could be due to reprocessing of the primary emission from a region of optically thick material: $L_{\rm BB}$ = ($\Omega$/4$\pi$) $L_{\rm X}$, where $\Omega$ is the solid angle subtended by the reprocessing material at a distance $R$ from the central X-ray source. If we assume that $L_{\rm BB} = \Omega R^2 \sigma T^4_{\rm BB}$, the distance $R$ can be estimated from the relation $R^2 = L_{\rm X}$/($4\pi \sigma T^4_{\rm BB}$). In the case of \smc, the total luminosity $L_{\rm X} =  1.4 \times 10^{38}$ \lum and the BB temperature $kT_{\rm BB}$ = 135 eV imply a distance $R \simeq 1.8 \times 10^8$ cm. If the reprocessing region is a shell at the inner edge of the accretion disc, $R$ should be of the order of the magnetospheric radius $R_{\rm m} \sim 1.5 \times 10^8 m_1^{1/7} R_6^{10/7} L_{37}^{-2/7} B_{12}^{4/7}$ cm, where $m_1$ is the NS mass in units of solar masses, $R_6$ is the NS radius in units of $10^6$ cm, $L_{37}$ is the X-ray luminosity in units of $10^{37}$ \lum, and $B_{12}$ is the NS magnetic field in units $10^{12}$ G \citep{Davies&Pringle81}. Assuming $m_1$ = 1.4, $R_6$ = 1 and $B_{12}$ = 1, for \smc\ we obtain $R_{\rm m} \simeq 10^8$ cm, comparable to $R$. We found that a reliable description of the soft component can be obtained also with a hot thermal plasma model (\textsc{apec} in \textsc{xspec}), able to account also for the blend of lines at $\sim$ 1 keV, without any Gaussian component. If it is left free to vary, the best-fit metal abundance (0.034$^{+0.015}_{-0.014}$) is significantly lower than that estimated for the SMC (0.2), but we verified that an acceptable fit can be obtained also with an abundance fixed at 0.2.

The RGS spectra show several structures: narrow emission lines due to ionized N, O, Ne, and Si; a broad feature at $\sim$ 1 keV (detected also with EPIC), which can be due to a blend of Fe-L lines or to RRC from O {\sc viii} and Ne {\sc ix}. Moreover, a broad emission feature is detected at $\sim$ 6.6 keV; it has an Equivalent Width of $\sim$ 0.1 keV and can be attributed to K-shell emission from iron at various ionization levels. A Fe-K emission line was already detected by \ASCA\ in 2000, but with a lower energy (6.3 keV) and a much larger EW ($\sim$ 0.4 keV). Since an accretion disk is likely present to fuel the high accretion rate onto the pulsar, photoionized circumsource matter from the inner disk atmosphere or the companion wind can be potential sources for line emission. Photoionization is also supported by the predominance of the forbidden line ({\sc O vii} (f) in Table~3) in the He-like {\sc O vii} triplet, if our identification of the emission line at 557$^{+5}_{-4}$~eV with {\sc O vii} (f) is correct \citep{Liedahl2001}. From this point of view, we note that large residuals, corresponding to the narrow emission features, can be still observed in the RGS spectra when its continuum component is described with a \textsc{cutoffpl+apec} model. Since the \textsc{apec} component cannot account for these features, we favour photoionized circumsource matter instead of a thermal plasma as the origin of the observed lines.
\vspace{-0.5 cm}



\section*{Acknowledgments}
We acknowledge financial contribution from the agreement ASI-INAF I/037/12/0. NLP and LS acknowledge the grant from PRIN-INAF 2014 `Towards a unified picture of accretion in HMXRBs'.
\vspace{-1 cm}

\bibliographystyle{mn2e} 
\bibliographystyle{mnras}
\bibliography{biblio}

\bsp

\label{lastpage}

\end{document}